\documentclass[prl,
 reprint,
superscriptaddress,
preprintnumbers,
nofootinbib,
 amsmath,amssymb,
 aps,
]{revtex4-1}
\usepackage[ocgcolorlinks]{hyperref}
\usepackage{graphicx}

\usepackage[utf8]{inputenc}

\newcommand{\be}{\begin{equation}}
\newcommand{\ee}{\end{equation}}
\newcommand{\ba}{\begin{eqnarray}}
\newcommand{\ea}{\end{eqnarray}}

\renewcommand{\l}{\left(}
\renewcommand{\r}{\right)}
\newcommand{\la}{\langle}
\newcommand{\ra}{\rangle}
\newcommand{\e}{\mathrm{e}}

\newcommand{\half}{\frac{1}{2}}

\begin{document}

\preprint{INR-TH-2023-003}

\title{
Scalar decay into pions via Higgs portal}

\author{Dmitry Gorbunov}
\email{gorby@ms2.inr.ac.ru}
\affiliation{Institute for Nuclear Research of the Russian Academy of Sciences, 117312 Moscow, Russia}
\affiliation{Moscow Institute of Physics and Technology, 141700 Dolgoprudny, Russia}
\author{Ekaterina Kriukova}
\email{kryukova.ea15@physics.msu.ru}
\affiliation{Institute for Nuclear Research of the Russian Academy of Sciences, 117312 Moscow, Russia}
\affiliation{Lomonosov Moscow State University, 119991 Moscow, Russia}
\author{Oleg Teryaev}
\email{teryaev@theor.jinr.ru}
\affiliation{Lomonosov Moscow State University, 119991 Moscow, Russia}
\affiliation{Bogoliubov Laboratory of Theoretical Physics, Joint Institute for Nuclear Research, 141980 Dubna, Russia}

\date{\today}

\begin{abstract}
 In extensions of the Standard Model (SM) of particle physics a light scalar from a hidden sector can interact with known particles via mixing with the SM Higgs boson. If the scalar mass is of GeV scale, this coupling induces the scalar decay into light hadrons, that saturates the scalar width. 
 Searches for the light scalars are performed in many ongoing experiments and planned for the next generation projects. 
 Applying dispersion relations changes the leading order estimate of the scalar decay rate into pions by a factor of about a hundred indicating the strong final state interaction. This subtlety
 for about thirty years 
  prevented any reliable inference of the model parameters from experimental data. In this Letter we use the gravitational form factor for neutral pion extracted from analysis of $\gamma^*\gamma\to\pi^0\pi^0$ processes to estimate the quark contribution to scalar decay into two pions. We find a factor of two uncertainty in this estimate and argue that the possible gluon contribution is of the same order. The decay rate to pions smoothly matches that to gluons dominating for heavier scalars.  
With this finding we 
  refine sensitivities of future projects to the scalar-Higgs mixing. 
  The accuracy in the calculations can be further improved by performing similar analysis of $\gamma^*\gamma\to K K$ and $\gamma^*\gamma\to\eta\eta$ processes and possibly decays like $J/\psi\to\gamma+\pi\pi$. 
\end{abstract}

\maketitle

{\it 1.} 
New physics required to address neutrino oscillations, baryon asymmetry of the Universe, dark matter and other phenomena unexplained within the Standard Model of particle physics, can be confined in a hidden sector, so that the new particles are sterile  with respect to the SM gauge interactions. They still can couple to the SM particles not only via gravity. There can be interactions via contact terms constructed by specific field products, invariant under  the SM and hidden gauge groups. 

One of the intriguing examples follows from the so-called scalar Higgs-field portal \cite{Patt:2006fw}, which combines the SM Higgs field $H$ and a scalar $S$, singlet with respect to the SM gauge group, into the interaction 
\begin{equation}
\label{portal}
  {\cal L}=\mu S H^\dagger H + \lambda S^2 H^\dagger H\,.   
\end{equation}
When the SM Higgs field gets non-zero vacuum expectation value $v=246$\,GeV, the first term in eq.\,\eqref{portal} yields mixing between the scalar and the SM Higgs boson $h$. Note, that if $S$ is charged under the hidden sector gauge group, this term is absent. However, the mixing can still arise, if the hidden sector gauge group is spontaneously broken, similar to the electroweak gauge group of the SM. In this case the mixing between the Higgs boson and its analog in the hidden sector comes from the second term in\,\eqref{portal}. Without loss of generality in both cases the induced interaction between the SM Higgs boson $h$ and the hidden scalar $S$ can be described as the mixing mass term in the scalar sector 
\begin{equation}
\label{lagrangian}
{\cal L}_s=\half m_h^2h^2+\mu v S h+\half M_S^2 S^2\,. 
\end{equation}
Thus, if kinematically allowed, the hidden scalar can be produced in scatterings and decays of SM particles and can decay into the SM particles through the virtual Higgs boson provided $\mu\neq0$. 

Hereafter we are interested in the models, where the hidden scalar is lighter than the Higgs boson. This case may naturally be favored \cite{Vissani:1997ys,deGouvea:2014xba}, because the heavy scalars coupled to the SM Higgs field would induce large quantum corrections to its mass. Moreover, we concentrate on the situation, where the scalar is at a GeV mass scale and so it can be produced in particle collisions, including accelerator experiments. While this choice may look ad hoc, there are particular extensions of the SM which actually predict new scalars in this mass range, see e.g.\,\cite{Bezrukov:2009yw,Chen:2015vqy,Dev:2017dui}. 

Searches for such light scalars have been performed in beam-dump experiments, accelerator experiments on neutrino oscillations, collider experiments, precision measurements, hunting for rare processes, etc, see Ref.\,\cite{Batell:2022dpx} for the most recent summary. So far, the negative results imposed constraints on the scalar production and decay rates, which could not be reliably transferred to limits on the model parameters because of a factor of hundred difference in estimates of the scalar decay rates into light hadrons \cite{Donoghue:1990xh,Bezrukov:2009yw,Winkler:2018qyg} obtained for the scalar mass in the region of 1\,GeV within the dispersion relation approach, on one hand, and with extrapolation of the results of Chiral Perturbation Theory (ChPT) and Quantum Chromodynamics (QCD), on the other hand. The two extrapolations give similar results, and 
while neither ChPT nor QCD application is justified in this mass region, the dispersion relation approach in this case can be also 
questioned\,\cite{Clarke:2013aya,Monin:2018lee,Bezrukov:2018yvd}. Hence, any  alternative way to tackle the problem is welcome. 

In this {\it Letter} we perform the calculation of the scalar decay rate into a couple of pions with the help of Generalized Distribution Amplitudes\,\cite{Diehl:1998dk,Diehl:2003ny}. The obtained result happens to be much closer to the naive extrapolations of the both ChPT and QCD results to the region of 1\,GeV in comparison with what the dispersion relation approach gives. 
This may be considered as an example of matching twist and chiral extrapolations which appeared 
to be rather effective in description of generalized Gerasimov-Drell-Hearn sum rules \cite {Soffer:2004ip}. 
We argue that the  uncertainty of our result is about a factor of two and present the refined estimates of the sensitivity of future experiments to the model parameters.

To begin with, we note that the Lagrangian \eqref{lagrangian} can be diagonalized. The resulting  scalar couplings to the SM fields are those of the SM Higgs boson couplings multiplied by the corresponding mixing angle $\xi$. The latter is strongly constrained by negative results of the experimental searches, $\xi\ll 1$. In this regime we can safely use the same notations $S$, $h$ and names for the true mass states in the scalar sector. The scalar effective interaction with quarks $q$ and gluons $g$ is described by the Lagrangian   
\begin{equation}
    \label{QCD}
{\cal L}_{qg}=-\xi \,S \sum_q \frac{m_q}{v}\bar q q + \xi\,S\frac{\alpha_s\,N_h}{12\pi\,v} G_{\mu\nu}^a G^{\mu\nu\,a}\,,
\end{equation}
where $m_q$ are quark masses. The first term in \eqref{QCD} is from the Yukawa couplings of the SM Higgs boson. Then $N_h$ heavy quarks,  i.e. $m_q\gg M_S/2$, induce the second term by quantum corrections, there $G_{\mu\nu}^a$ is gluonic field tensor, $a=1,\dots,8$, and strong coupling  $\alpha_s$ (being the QCD analogue of the fine structure constant)  is evaluated at the scale of the order of $M_S$. This Lagrangian allows one to estimate the scalar decay rates to quarks, 
\begin{equation}
    \label{to-quarks}
    \Gamma(S\to\bar q q)=\xi^2\frac{N_c}{8\pi}\frac{m_q^2\,M_S}{v^2}\l 1-\frac{4\,m_q^2}{M_S^2}\r^{3/2}
\end{equation}
(where $N_c=3$ is the number of quark color states) and to gluons,  
\begin{equation}
    \label{to-gluons}
    \Gamma(S\to gg)=\xi^2\frac{N_c^2-1}{8}\frac{N_h^2\alpha_s^2}{72\pi^3}\frac{M_S^3}{v^2}
\end{equation}
(where $N_c^2-1=8$ is the number of gluon states). 
Numerically, quark modes dominate over the gluon mode for heavy scalars. However, at GeV scale only $u$, $d$ and $s$ quarks are relevant, $N_h=3$, and gluons become important. 

Light scalars decay directly into meson pairs, and these hadronic decay rates can be described directly via effective interaction between the scalar and light mesons. It can be obtained by making use of the renorminvariance of the hadronic contribution to the trace of the energy-momentum tensor \cite{Vainshtein:1980ea},   
\begin{equation}
\label{trace}
T^\mu_\mu\equiv \sum_{q=u,d,s} m_q\bar q q - 
\frac{9\alpha_s}{8\pi} G_{\mu\nu}^a G^{\mu\nu\,a}\,,
\end{equation}
which we present to the leading order in $\alpha_s$. The last term in \eqref{trace} comes from violation of the scale invariance by the trace anomaly due to the running of $\alpha_s$ with energy. It is generated by one-loop triangle diagrams with virtual $u,d,s$ quarks,  and for heavy quarks the contributions of two terms cancel. This relation allows one to recast the light scalar interaction \eqref{QCD} in terms of quarks and $T^\mu_\mu$ as 
\begin{equation}
    \label{quarks-trace}
{\cal L}_{T}=-\xi \,\frac{S}{v} \l \l 1-\frac{2N_h}{27}\r\!\!\!\sum_{q=u,d,s} m_q\bar q q + \frac{2\,N_h}{27} T^\mu_\mu\r. 
\end{equation}
Therefore, to calculate the scalar decay rates to, say, pions, one must evaluate the matrix elements ($a,b=1,2,3$)
\begin{align}
\label{GammaP}
  \la \pi^a(p)\pi^b(p')| m_u\bar u u + m_d \bar d d| 0\ra &\equiv \delta^{ab} \Gamma_\pi (s)\,,  \\
\label{DeltaP}
\la \pi^a(p)\pi^b(p')| m_s\bar s s | 0\ra &\equiv \delta^{ab} \Delta_\pi (s)\,, \\
\label{TP}
\la \pi^a(p)\pi^b(p')| T^\mu_\mu | 0\ra &\equiv \delta^{ab} T_\pi (s)
\end{align}
and the similar elements for decays into kaons, $\eta$-mesons, etc. 

The form factors entering eqs.\,\eqref{GammaP}-\eqref{TP} depend on the invariant mass of pion states, $s=(p+p')^2$. They can be calculated within the Chiral Perturbation Theory (ChPT). The leading order terms read
\begin{align}
\label{GammaP-ChPT}
 \Gamma_\pi (s) & = m_\pi^2\,,  \\
\label{DeltaP-ChPT}
\Delta_\pi (s) & = 0\,,\\
\label{TP-ChPT}
T_\pi (s) & = s + 2m_\pi^2\,,
\end{align}
where $m_\pi$ stands for the pion mass. Hence the amplitude of the scalar decay into pions is proportional to \cite{Voloshin:1985tc}
\begin{align}
    \label{total-amplitude-1}
    G_\pi(s=M_S^2)&\equiv  2\, T_\pi (s)+ 7\, \Gamma_\pi(s) +7\,\Delta_\pi(s) \\ & =11\,m_\pi^2+2 M_S^2\,. 
\label{total-amplitude-2}
\end{align}
However, adopting these formulas for evaluation of the decay rates of scalars as heavy as 1\,GeV can not be justified. Indeed, it was argued to be unreliable\,\cite{Donoghue:1990xh} 
due to the strong interaction of pions in the final states. The arguments were based on the usage of dispersion relations and extracted from $\pi\pi\to\pi\pi$ data $S$-matrix elements. The inferred corrections strongly depend on $s$ and change the leading-order estimate of the corresponding hadronic decay rate by a factor upto a 
hundred\,\cite{Winkler:2018qyg}. 

Later, the usage of dispersion relations in that case, limited by associated experimental uncertainties, adopted truncated $S$-matrix, etc, has been questioned in literature, 
e.g.\,\cite{Clarke:2013aya,Monin:2018lee,Bezrukov:2018yvd}, but no credible alternative estimate of the hadronic decay rates were presented.

{\it 2.} In this Letter we make use of the $q=u,d$ quark contributions to the gravitational \cite{Burkert:2023wzr,Teryaev:2016edw} form factors of a pion, considered \cite{Kumano:2017lhr} in the timelike domain. 
They are defined via {\it quark energy-momentum tensor} as  
\begin{equation}
    \label{definition}
    \begin{split}
&\la \pi^a(p)\pi^b(p')|T^{\mu\nu}_q(0)|0\ra 
\\ &\equiv \frac{\delta^{ab}}{2} \l \l s \,\eta^{\mu\nu}-P^\mu P^\nu\r \Theta_{1,q}(s)+\Delta ^\mu \Delta^\nu \Theta_{2,q}(s)\r, 
\end{split}
\end{equation}
where $P\equiv p+p'$ and $\Delta\equiv p'-p$. Convolution of \eqref{definition} with metric $\eta_{\mu\nu}$ and summation over quarks, $\Theta_i\equiv \Theta_{i,u}+\Theta_{i,d}$, gives for the quark form factor \eqref{GammaP} 
\begin{equation}
\label{GammaP-Theta}
\Gamma_\pi(s)=s \l \frac{3}{2}\Theta_{1}(s)-\half \Theta_{2}(s)\r +2 m^2_\pi \Theta_{2}(s)\,.
\end{equation}
There is also the contribution to the quark part of $T_\mu^\mu$ in \eqref{trace} and hence in \eqref{TP}, which we take into account in due course.

The form factors $\Theta_{1(2),q}(s)$ have been inferred by fitting the experimental data from Belle on $\gamma ^* \gamma \to \pi^0\pi^0$ scattering within the technique of Generalized Distribution Amplitudes\,\cite{Diehl:1998dk,Diehl:2003ny}. The fitting formulas read (we use the original notations from \cite{Kumano:2017lhr} correcting the obvious typo: extra $\beta^2$-factor in the resonance term in $\tilde B_{20}$): 
\begin{align}
    \Theta_{1,q}(s)&=-\frac{3}{5}\tilde B_{10}(s)+\frac{3}{10}\tilde B_{20}(s) \\
    \Theta_{2,q}(s)&=\frac{9}{10\beta^2}\tilde B_{20}(s)
\end{align}
where $\beta^2=\beta^2(s)\equiv 1-4m_\pi^2/s$ \; and 
\begin{align*}
\tilde B_{10}(s)=&-\frac{10}{9} \left[ \l 1+\frac{2m_\pi^2}{s}\r 
M_{2(q)}^\pi F_q^\pi(s) \right. \\ & \left. +\frac{3 g_{f_0\pi\pi}\bar f_{f_0}}{2\sqrt{2} 
\sqrt{ (M_{f_0}^2-s)^2 + \Gamma_{f_0}^2 M_{f_0}^2}} \right] 
\e^{i\delta_0(\sqrt{s})},\\
\tilde B_{20}(s) &=  \frac{10}{9}\beta^2 \left[ M_{2(q)}^\pi  F_q^\pi(s) \right. \\
& + \left.
\frac{g_{f_2\pi\pi}f_{f_2} M^2_{f_2}}{\sqrt{2} \sqrt{(M_{f_2}^2-s)^2 + \Gamma_{f_2}^2 M_{f_2}^2}} \right] 
\e^{i\delta_2(\sqrt{s})}
\end{align*}
with $F_q^\pi (s)=(1+\beta^2s/\Lambda^2)^{-1}$
and the relative contribution of quarks to the total pion momentum  
$M^\pi_{2(u)}+M_{2(d)}^\pi=0.5$.  
The phase shifts for $S$ and $D$-waves were taken from numerical fit of Ref.\,\cite{Bydzovsky:2016vdx}, and the first one was corrected above the kaon threshold as $ 
\delta_0(\sqrt{s})\to\delta_0(\sqrt{s})+a_\delta \l (\sqrt{s}-2m_K)/1\,\text{GeV} \r^{b_\delta}\,.
$
The values (at the scattering energy scale) of the resonance parameters used in the fit are presented in Tab.\,\ref{tab:resonances}.
\begin{table}[!htb]
    \centering
    \begin{tabular}{|c|c|c|c|c|}
\hline Meson ($h$) & $M_h$\;(GeV) & $\Gamma_h$\;(GeV) & $g_{h\pi\pi}$        &  $f_h$ (GeV)\\
  \hline   
  $f_0$(500)   &   0.475 &  0.550 & 2.959\,GeV & -- 
    \\
  \hline   
  $f_2$(1270)   &   1.275 &  0.185 & 1.953\,GeV$^{-1}$ & 0.0754\\
  \hline
\end{tabular}
    \caption{Parameters of the hadronic resonances entering the fitting formulas for $\tilde B_{i0}$.}
    \label{tab:resonances}
\end{table}
These numbers coincide with those in Tab.I of Ref.\,\cite{Kumano:2017lhr} after  
correcting the typo in the value of $g_{f_2\pi\pi}$ 
which was presented there being smaller by a factor of 12.44 (however not affecting the code and final results).       

\begin{figure}[!htb]
    \centering
    \includegraphics[width=\columnwidth]{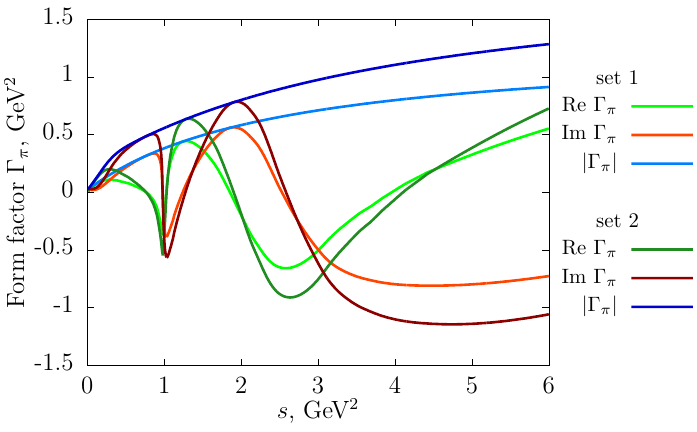}
    \vspace*{-6mm}
    \caption{Real, imaginary parts and the absolute value of $\Gamma_\pi$ for the fitting sets 1 and 2 of Tab.\,\ref{tab:parameters}.}
    \label{fig:Gamma-Pi}
    \vspace*{-2mm}
\end{figure}

Numerical fit to the Belle data on $\gamma^*\gamma\to\pi^0\pi^0$ revealed the values of the fitting parameters summarized in Tab.\,\ref{tab:parameters}.
\begin{table}[!htb]
    \centering
    \begin{tabular}{|c|c|c|}
\hline
Parameter         &   set 1 & set 2\\\hline 
$\alpha$  & $0.801 \pm 0.042$ &  $1.157\pm 0.132$ \\
$\Lambda$ (GeV) & $1.602 \pm 0.109$ & $1.928 \pm 0.213$ \\
$\bar f_{f_0(500)}$ (GeV) & 0 (fixed) & $0.0184\pm 0.0034$ \\
$a_\delta$ & $3.878\pm 0.165$ & $3.800\pm 0.170$ \\
$b_\delta$ & $0.382\pm 0.040$ & $0.407 \pm 0.041$ \\ 
\hline 
$\chi^2/$d.o.f. & 1.22 & 1.09 \\ \hline
    \end{tabular}
    \caption{Sets of fitting parameters adopted from Ref.\,\cite{Kumano:2017lhr}.}
    \label{tab:parameters}
\end{table}
With formulas and parameters above we accurately restore the real and imaginary parts of the form factors $\Theta_{i,q}$ presented in Fig. 19 of \cite{Kumano:2017lhr}. 
There are two sets 
with similar accuracy of the fitting, which provides an estimate of the uncertainty in the gravitational form factors, and hence in the scalar decay rate to pions, associated with the exploited methods and experimental data. Numerical results for real, imaginary parts and the absolute value of $\Gamma_\pi$ for $\Theta_{i,q}$ determined by these  fits are shown in Fig.\,\ref{fig:Gamma-Pi}.
The fitting procedure of Ref.\,\cite{Kumano:2017lhr} adopts as the boundary condition at $s\to 0$ the leading order prediction of ChPT\,\eqref{GammaP-ChPT}.  With $x\equiv 
(M_S^2-4m_\pi^2)/(1\, \text{GeV}^2)$ we obtain a numerical approximation for the average between set 1 and set 2:
\begin{align}
\label{G-fit}
    &|\Gamma_\pi(x)|=(0.0687 + 0.0270 x^{1/2} + 0.697 x^{3/4} -\\ \nonumber
    & -0.283 x)\text{ GeV}^2.
\end{align}

\begin{figure}[!htb]
    \centering
    \includegraphics[width=\columnwidth]{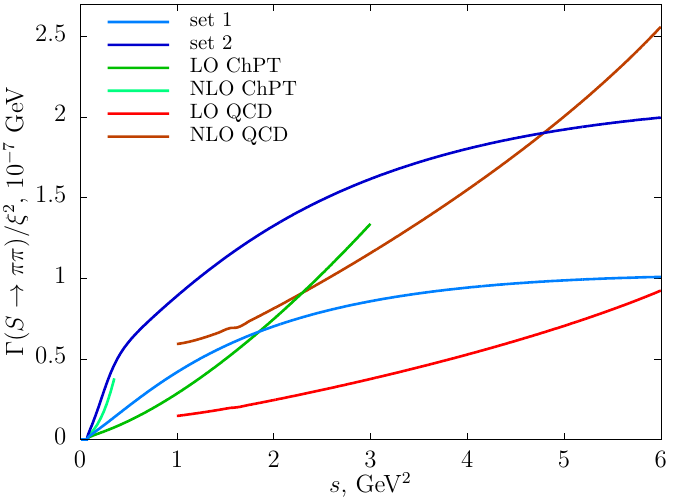}
    \vspace*{-6mm}
    \caption{The decay width of hidden scalar to pions calculated using $\Gamma_\pi$ from Fig.\,\ref{fig:Gamma-Pi} for the fitting sets 1 and 2 (light and dark blue lines) and ChPT (green line) divided by $\xi^2$. Light green line shows the decay width to pions divided by $\xi^2$ obtained using the NLO ChPT results for $\Gamma_\pi$ and $T_\pi$ \cite{Gasser:1983yg,Kubis:1999db}. The result of LO (NLO \cite{Djouadi:1991tka}) QCD calculation for the decay width of the hidden scalar to gluons is shown in red (dark orange), the renormalization scale is $\sqrt{s} \equiv M_S$.}
    \label{fig:Decay-width}
    \vspace*{-2mm}
\end{figure}

{\it 3.} 
The estimated form factor contributes to the total scalar decay rate to pions (the rate to neutral pions is half of the rate to charged ones) as 
\begin{align}
\label{rate}
    &\Gamma(S\to \pi\pi)=\frac{3}{32\pi}\frac{49\xi^2 \, |\Gamma_\pi(M_S^2)|^2}{81\,v^2M_S}\,\beta(M_S^2)\\ \nonumber
    &= (0.983+6.54y-1.12y^2+0.071y^3)\cdot 10^{-8}\text{ GeV}\,,
\end{align}
where $y\equiv M_S^2/1$\,GeV$^2$,
while the leading order ChPT result is obtained from \eqref{rate} 
by replacing $\Gamma_\pi\to G_\pi$, see \eqref{total-amplitude-1}. Both results are outlined in Fig.\,\ref{fig:Decay-width}
along with the leading-order QCD calculations of the decay rate into gluons\,\eqref{to-gluons} and the next-to-leading order ones calculated for the light Higgs boson at the renormalization scale $\mu=M_S$ \cite{Djouadi:1991tka}. There is also shown the decay width obtained using the NLO ChPT result for $\Gamma_\pi$ and $T_\pi$ \cite{Gasser:1983yg,Kubis:1999db}. 
We observe, that with our estimate of $\Gamma_\pi$ the decay rate to pions reasonably matches that into gluons at $M_S\simeq$\,1.5-2\,GeV (light quark contribution \eqref{to-quarks} is negligible), and in this mass range the total ChPT leading order estimate reveals similar result.
One may therefore speak on "gluon-hadron" (or in some sense "quark-gluon", as the pions are obviously quark states) duality. While the smaller fitting curve seems to be preferable, 
the deviations by no means may substantially exceed a factor of two, still consisting with uncertainties we expect in the fitting and calculations. 
One should bear in mind that the ChPT expansion is valid only up to dark scalar masses of order 1 GeV, although we extend the green line in Fig.\,\ref{fig:Decay-width} to higher $s$ in order to demonstrate the overlapping of the results. In this region also the power corrections should be 
important and their recent studies \cite{Lorce:2022tiq} are in line with the estimate of uncertainties mentioned above. 

One may consider the closeness of the curves for NLO ChPT and NLO QCD as a manifestation of possibility to match the expansions in positive powers of $Q^2$ in ChPT and that in $log Q^2$ in NLO QCD supplemented by negative powers of 
$Q^2$ in higher twist \cite{Lorce:2022tiq} corrections. This may, in turn, be a sort of a sign that the region of $Q^2$ 
in which neither ChPT nor NLOQCD+HT expansions are valid is relatively narrow. 
The additional closeness to the third curve (set 1) in this region is supporting this assumption.
As it was already mentioned, this situation is also similar to the description of spin-dependent DIS, where matching  \cite {Soffer:2004ip} of different expansions 
provides rather good description of the data. 

Note that the scalar form factor behaviour 
within dispersion approach and ChPT were systematically compared in Sec.\,2 of \cite{Colangelo:2001uv}. 
While both NNLO contribution of ChPT and the dispersion relation provide a decrease of $\Gamma_\pi(s)$ 
at $s \sim (0.5\,\textrm{GeV})^2$, there is a  quantitative discrepancy due to an underestimate of phase in ChPT. Moreover, at $s \sim 1\,\textrm{GeV}^2$ the phase evaluation in the Omnes approach should be strongly violated by the contributions of inelastic channels, in particular, $KK$,
making their account rather important. Indeed $\Gamma_\pi$ exhibits much lower peak within 2-channels analysis \cite{Bezrukov:2018yvd}. 

Contrary to estimates of\,\cite{Donoghue:1990xh}, our result for the decay rate as a function of scalar mass does not exhibit any peak-like structures, which might be attributed to the impacts of light scalar hadronic resonances. We find that impacts of some, in particular $f_0(500)$ and $f_2(1270)$, are small, while some, e.g. $f_0(980)$, are not seen in the fit \cite{Kumano:2017lhr} and hence  do not interfere in \eqref{GammaP} \emph{being most probably bound states of four quarks},  see e.g.\,\cite{Achasov:2020aun,ParticleDataGroup:2022pth}. 
Meanwhile, the latter resonance is the main source of the peak-like feature at about 1\,GeV$^2$ in the shape of $\Gamma_\pi(s)$ presented in Ref.\,\cite{Donoghue:1990xh}. It is responsible for a factor of hundred deviation of our result for the scalar decay rate into pion from that of Ref.\,\cite{Donoghue:1990xh}.

Our approach treats the scalar decay to pion pair induced by the bilinear quark operator as production of a quark-antiquark  pair which goes to a couple of pions via all possible virtual hadronic states. Naturally, $f_0(980)$ does not substantially contribute to this amplitude, which can be sketched as 
\begin{equation}
\la S|\pi\pi\ra \to \la S|\e^{i\int S \bar q q d^4x}|{\rm hadrons}\ra \la {\rm hadrons} | \pi\pi\ra.
\label{compl}
\end{equation}
On the contrary, the dispersion relation method implies that the scalar decays exclusively to a pion pair, which rescatter, that is described as a pion pair transformed into a pion pair via all possible hadronic states, i.e. 
\begin{eqnarray}
\la S|\pi\pi\ra \to \la S|\e^{i\int S \bar q q d^4x}|\pi\pi\ra\la\pi\pi|{\rm hadrons}\ra \ 
 \nonumber \\ \times \la {\rm hadrons} | \pi\pi\ra.
\label{compll}
\end{eqnarray}
In that case $f_0(980)$ may be present among the intermediate hadronic states. 

One may also refer to the notion of quark-hadron duality for energy-momentum tensor. It is bilinear in Dirac fields at the quark level and bilinear in meson fields at the hadronic one. This may provide the qualitative explanation of the observed decoupling of four-quark states from the amplitude of two-meson decays controlled by the energy-momentum tensor.

While all the states are mixed in quantum field theory, the specifics of duality for the tetraquark states means the smallness of their coupling to bilinear currents, including the energy-momentum tensor. This smallness may be considered as a complementary ("duality") way to express the known fact \cite{Achasov:2020aun,ParticleDataGroup:2022pth} of four-quark nature  of $f_0(980)$. 
It is instructive to compare the duality arguments with representations (\ref{compl},\ref{compll}). The duality    means the complementarity between different choices of full set of (quark or hadronic) intermediate states. This full set appears in the imaginary part of amplitude and after
recovering the real part via the dispersion relations results in some kind of 
sum rules.  
At the same time, the projection onto pion states in \eqref{compll} when the final state interaction is taken into account via the Omn\'es method (requiring to restrict the 
set of states involved) in the dispersion relations 
framework, may violate the complementarity between quark and hadronic description.   

One may expect that $f_0(980)$ will be more pronounced for  the case of multi-quark operators. They control, in particular, the higher twist contributions responsible for power corrections which will be more pronounced at lower $Q^2$ region. The real photon $Q^2 \to 0$ limit should involve the infinite tower of higher twists describing the two-photon decay width of $f_0(980)$.

This situation is somehow similar to the exclusive  production of $\rho$ meson pairs \cite{Anikin:2005ur}, when the role of momentum conservation is taken by that of  the isospin. As the quark-antiquark pair cannot have isospin $I=2$, only the $I=0$ meson states are produced at large $Q^2$. At the same time, for real photons the $I=2$ four-quark state contributes \cite{Achasov:1982bt}, yielding a dramatic change in the ratio of charged and neutral mesons production rates.

Note in passing that operator $m\bar q q$ contributes also to $T_\pi$. Its account implies the replacement $49/81\to 1$ in eq.\,\eqref{rate}.

\begin{figure}[!htb]
    \centering
    \includegraphics[width=\columnwidth]{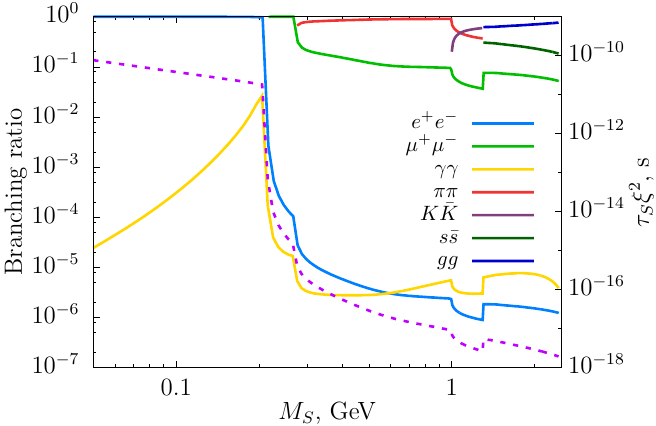}
    \vspace*{-6mm}
    \caption{Branching ratios of hidden scalar to leptons, photon pairs, pions (using the result of this work), kaons (within the framework of ChPT), $s\bar{s}$ pairs and gluons (NLO QCD calculation). The branchings are shown with solid lines. The dashed line shows hidden scalar lifetime multiplied by $\xi^2$.}
    \label{fig:Branching}
    \vspace*{-2mm}
\end{figure}

{\it 4.} 
The calculated decay rate of the hidden scalar into pions 
can be used along the decay rates into photons, leptons, etc, see e.g.\,\cite{Bezrukov:2009yw}, to evaluate the light scalar lifetime and entire pattern of branching ratios of the light scalar decays into the SM particles, see Fig.\,\ref{fig:Branching}.
Here for $\Gamma_\pi(s)$ we use \eqref{G-fit}. In the mass region $M_S\sim 1$\,GeV we expect uncertainties by a factor of 2-3 due to uncertainties in the gravitational form factor we used (the difference between two fits in Fig.\,\ref{fig:Decay-width}, where different resonances have been accounted for, illustrates the size of these uncertainties inherent in the approach) and due to our disregard of the gluonic contribution to the latter. It might be partially accounted along the quark contribution in the analysis of Ref.\,\cite{Kumano:2017lhr}, which deserves an elaboration.
At the moment, the gluonic effects may be estimated by the size of pion momentum fraction carried by gluons being of about 
$30\%$ \cite{Barry:2018ort}.

\begin{figure}[!htb]
    \centering
    \includegraphics[width=\columnwidth]{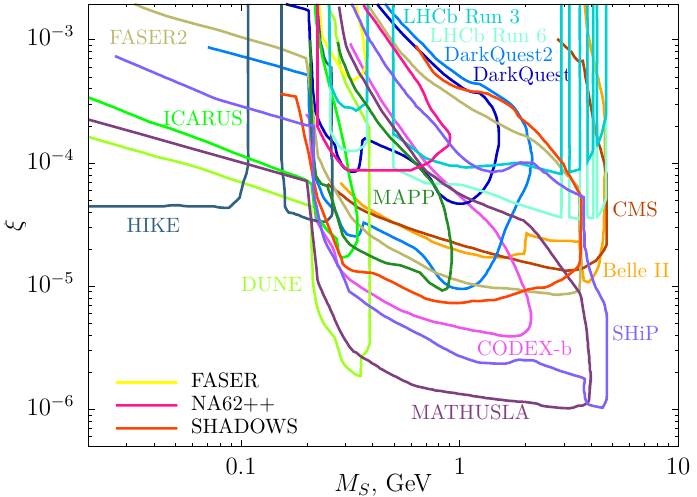}
    \vspace*{-6mm}
    \caption{Expected reach of existing and proposed experiments in the model parameter space updated in accordance with the result for decay rate to pions. Peaks of original curves from \cite{Batell:2022dpx} corresponding to the resonance enhancement of $\Gamma(S\to \pi\pi)$ have been smoothed accordingly.}
    \label{fig:Exp}
    \vspace*{-2mm}
\end{figure}

 Assuming that fit parameters from Tab.\,\ref{tab:parameters} are normally distributed, the relative $1\sigma$ uncertainty in dark scalar decay rate to pions can be estimated as 4-19$\%$ for set 1 and 10-17$\%$ for set 2 depending on $s$. The total uncertainties are not expected to be large, given the nice match of our result to that of QCD in the region of $s=4\!-\!6$\,GeV$^2$. We do not expect any induced by gluons features in the scalar decay rate to hadrons provided no light scalar resonances consisted of gluons. We use the corrected pattern in Fig.\,\ref{fig:Branching} to refine the experimental reach in the model parameter space as presented in Fig.\,\ref{fig:Exp}. 
The new sensitivity curves are smooth, while the original curves exhibited features at the scalar mass of about 1\,GeV. The corresponding change in $\xi$ is about 3-10, depending on the signatures adopted for the projected searches for the hidden scalar.

In this paper we put forward a new method to calculate the hadronic decay rates of a hypothetical ${\cal O}(1)$\,GeV-mass scalar coupled to the SM via Higgs-portal\,\eqref{portal}. Further work is needed to  estimate quantitatively the uncertainty of the calculations: what we did here is largely based on the phenomenological grounds. Note that the application of the dispersion relation approach to calculation of the scalar decay rates to light hadrons would benefit from similar study as well.

To improve the accuracy in calculation of the scalar decay rates to hadrons within our approach, it is worth to infer other hadronic gravitational form factors, possibly from analyses of $\gamma^*\gamma\to KK$, $\gamma^*\gamma\to\eta\eta$ scatterings or $J/\psi\to\gamma+\pi\pi$, etc, decays collected at $c$- and $b$-factories.



\begin{acknowledgments}
We thank I.\,Timiryasov for stimulating discussions. OT is indebted to S. Kumano, B. Pire and Qin-Tao Song for helpful correspondence. The work is partially supported by the  
Russian Science Foundation RSF grant 21-12-00379. The work of EK is supported by the grant of ``BASIS" Foundation no. 21-2-10-37-1.
\end{acknowledgments}

\bibliography{refs}

\begin{thebibliography}{31}%
\makeatletter
\providecommand \@ifxundefined [1]{%
 \@ifx{#1\undefined}
}%
\providecommand \@ifnum [1]{%
 \ifnum #1\expandafter \@firstoftwo
 \else \expandafter \@secondoftwo
 \fi
}%
\providecommand \@ifx [1]{%
 \ifx #1\expandafter \@firstoftwo
 \else \expandafter \@secondoftwo
 \fi
}%
\providecommand \natexlab [1]{#1}%
\providecommand \enquote  [1]{``#1''}%
\providecommand \bibnamefont  [1]{#1}%
\providecommand \bibfnamefont [1]{#1}%
\providecommand \citenamefont [1]{#1}%
\providecommand \href@noop [0]{\@secondoftwo}%
\providecommand \href [0]{\begingroup \@sanitize@url \@href}%
\providecommand \@href[1]{\@@startlink{#1}\@@href}%
\providecommand \@@href[1]{\endgroup#1\@@endlink}%
\providecommand \@sanitize@url [0]{\catcode `\\12\catcode `\$12\catcode
  `\&12\catcode `\#12\catcode `\^12\catcode `\_12\catcode `\%12\relax}%
\providecommand \@@startlink[1]{}%
\providecommand \@@endlink[0]{}%
\providecommand \url  [0]{\begingroup\@sanitize@url \@url }%
\providecommand \@url [1]{\endgroup\@href {#1}{\urlprefix }}%
\providecommand \urlprefix  [0]{URL }%
\providecommand \Eprint [0]{\href }%
\providecommand \doibase [0]{http://dx.doi.org/}%
\providecommand \selectlanguage [0]{\@gobble}%
\providecommand \bibinfo  [0]{\@secondoftwo}%
\providecommand \bibfield  [0]{\@secondoftwo}%
\providecommand \translation [1]{[#1]}%
\providecommand \BibitemOpen [0]{}%
\providecommand \bibitemStop [0]{}%
\providecommand \bibitemNoStop [0]{.\EOS\space}%
\providecommand \EOS [0]{\spacefactor3000\relax}%
\providecommand \BibitemShut  [1]{\csname bibitem#1\endcsname}%
\let\auto@bib@innerbib\@empty
\bibitem [{\citenamefont {Patt}\ and\ \citenamefont
  {Wilczek}(2006)}]{Patt:2006fw}%
  \BibitemOpen
  \bibfield  {author} {\bibinfo {author} {\bibfnamefont {B.}~\bibnamefont
  {Patt}}\ and\ \bibinfo {author} {\bibfnamefont {F.}~\bibnamefont {Wilczek}},\
  }\href@noop {} {\  (\bibinfo {year} {2006})},\ \Eprint
  {http://arxiv.org/abs/hep-ph/0605188} {arXiv:hep-ph/0605188} \BibitemShut
  {NoStop}%
\bibitem [{\citenamefont {Vissani}(1998)}]{Vissani:1997ys}%
  \BibitemOpen
  \bibfield  {author} {\bibinfo {author} {\bibfnamefont {F.}~\bibnamefont
  {Vissani}},\ }\href {\doibase 10.1103/PhysRevD.57.7027} {\bibfield  {journal}
  {\bibinfo  {journal} {Phys. Rev. D}\ }\textbf {\bibinfo {volume} {57}},\
  \bibinfo {pages} {7027} (\bibinfo {year} {1998})},\ \Eprint
  {http://arxiv.org/abs/hep-ph/9709409} {arXiv:hep-ph/9709409} \BibitemShut
  {NoStop}%
\bibitem [{\citenamefont {de~Gouvea}\ \emph {et~al.}(2014)\citenamefont
  {de~Gouvea}, \citenamefont {Hernandez},\ and\ \citenamefont
  {Tait}}]{deGouvea:2014xba}%
  \BibitemOpen
  \bibfield  {author} {\bibinfo {author} {\bibfnamefont {A.}~\bibnamefont
  {de~Gouvea}}, \bibinfo {author} {\bibfnamefont {D.}~\bibnamefont
  {Hernandez}}, \ and\ \bibinfo {author} {\bibfnamefont {T.~M.~P.}\
  \bibnamefont {Tait}},\ }\href {\doibase 10.1103/PhysRevD.89.115005}
  {\bibfield  {journal} {\bibinfo  {journal} {Phys. Rev. D}\ }\textbf {\bibinfo
  {volume} {89}},\ \bibinfo {pages} {115005} (\bibinfo {year} {2014})},\
  \Eprint {http://arxiv.org/abs/1402.2658} {arXiv:1402.2658 [hep-ph]}
  \BibitemShut {NoStop}%
\bibitem [{\citenamefont {Bezrukov}\ and\ \citenamefont
  {Gorbunov}(2010)}]{Bezrukov:2009yw}%
  \BibitemOpen
  \bibfield  {author} {\bibinfo {author} {\bibfnamefont {F.}~\bibnamefont
  {Bezrukov}}\ and\ \bibinfo {author} {\bibfnamefont {D.}~\bibnamefont
  {Gorbunov}},\ }\href {\doibase 10.1007/JHEP05(2010)010} {\bibfield  {journal}
  {\bibinfo  {journal} {JHEP}\ }\textbf {\bibinfo {volume} {05}},\ \bibinfo
  {pages} {010} (\bibinfo {year} {2010})},\ \Eprint
  {http://arxiv.org/abs/0912.0390} {arXiv:0912.0390 [hep-ph]} \BibitemShut
  {NoStop}%
\bibitem [{\citenamefont {Chen}\ \emph {et~al.}(2016)\citenamefont {Chen},
  \citenamefont {Davoudiasl}, \citenamefont {Marciano},\ and\ \citenamefont
  {Zhang}}]{Chen:2015vqy}%
  \BibitemOpen
  \bibfield  {author} {\bibinfo {author} {\bibfnamefont {C.-Y.}\ \bibnamefont
  {Chen}}, \bibinfo {author} {\bibfnamefont {H.}~\bibnamefont {Davoudiasl}},
  \bibinfo {author} {\bibfnamefont {W.~J.}\ \bibnamefont {Marciano}}, \ and\
  \bibinfo {author} {\bibfnamefont {C.}~\bibnamefont {Zhang}},\ }\href
  {\doibase 10.1103/PhysRevD.93.035006} {\bibfield  {journal} {\bibinfo
  {journal} {Phys. Rev. D}\ }\textbf {\bibinfo {volume} {93}},\ \bibinfo
  {pages} {035006} (\bibinfo {year} {2016})},\ \Eprint
  {http://arxiv.org/abs/1511.04715} {arXiv:1511.04715 [hep-ph]} \BibitemShut
  {NoStop}%
\bibitem [{\citenamefont {Dev}\ \emph {et~al.}(2017)\citenamefont {Dev},
  \citenamefont {Mohapatra},\ and\ \citenamefont {Zhang}}]{Dev:2017dui}%
  \BibitemOpen
  \bibfield  {author} {\bibinfo {author} {\bibfnamefont {P.~S.~B.}\
  \bibnamefont {Dev}}, \bibinfo {author} {\bibfnamefont {R.~N.}\ \bibnamefont
  {Mohapatra}}, \ and\ \bibinfo {author} {\bibfnamefont {Y.}~\bibnamefont
  {Zhang}},\ }\href {\doibase 10.1016/j.nuclphysb.2017.07.021} {\bibfield
  {journal} {\bibinfo  {journal} {Nucl. Phys. B}\ }\textbf {\bibinfo {volume}
  {923}},\ \bibinfo {pages} {179} (\bibinfo {year} {2017})},\ \Eprint
  {http://arxiv.org/abs/1703.02471} {arXiv:1703.02471 [hep-ph]} \BibitemShut
  {NoStop}%
\bibitem [{\citenamefont {Batell}\ \emph {et~al.}(2022)\citenamefont {Batell},
  \citenamefont {Blinov}, \citenamefont {Hearty},\ and\ \citenamefont
  {McGehee}}]{Batell:2022dpx}%
  \BibitemOpen
  \bibfield  {author} {\bibinfo {author} {\bibfnamefont {B.}~\bibnamefont
  {Batell}}, \bibinfo {author} {\bibfnamefont {N.}~\bibnamefont {Blinov}},
  \bibinfo {author} {\bibfnamefont {C.}~\bibnamefont {Hearty}}, \ and\ \bibinfo
  {author} {\bibfnamefont {R.}~\bibnamefont {McGehee}},\ }in\ \href@noop {}
  {\emph {\bibinfo {booktitle} {{2022 Snowmass Summer Study}}}}\ (\bibinfo
  {year} {2022})\ \Eprint {http://arxiv.org/abs/2207.06905} {arXiv:2207.06905
  [hep-ph]} \BibitemShut {NoStop}%
\bibitem [{\citenamefont {Donoghue}\ \emph {et~al.}(1990)\citenamefont
  {Donoghue}, \citenamefont {Gasser},\ and\ \citenamefont
  {Leutwyler}}]{Donoghue:1990xh}%
  \BibitemOpen
  \bibfield  {author} {\bibinfo {author} {\bibfnamefont {J.~F.}\ \bibnamefont
  {Donoghue}}, \bibinfo {author} {\bibfnamefont {J.}~\bibnamefont {Gasser}}, \
  and\ \bibinfo {author} {\bibfnamefont {H.}~\bibnamefont {Leutwyler}},\ }\href
  {\doibase 10.1016/0550-3213(90)90474-R} {\bibfield  {journal} {\bibinfo
  {journal} {Nucl. Phys. B}\ }\textbf {\bibinfo {volume} {343}},\ \bibinfo
  {pages} {341} (\bibinfo {year} {1990})}\BibitemShut {NoStop}%
\bibitem [{\citenamefont {Winkler}(2019)}]{Winkler:2018qyg}%
  \BibitemOpen
  \bibfield  {author} {\bibinfo {author} {\bibfnamefont {M.~W.}\ \bibnamefont
  {Winkler}},\ }\href {\doibase 10.1103/PhysRevD.99.015018} {\bibfield
  {journal} {\bibinfo  {journal} {Phys. Rev. D}\ }\textbf {\bibinfo {volume}
  {99}},\ \bibinfo {pages} {015018} (\bibinfo {year} {2019})},\ \Eprint
  {http://arxiv.org/abs/1809.01876} {arXiv:1809.01876 [hep-ph]} \BibitemShut
  {NoStop}%
\bibitem [{\citenamefont {Clarke}\ \emph {et~al.}(2014)\citenamefont {Clarke},
  \citenamefont {Foot},\ and\ \citenamefont {Volkas}}]{Clarke:2013aya}%
  \BibitemOpen
  \bibfield  {author} {\bibinfo {author} {\bibfnamefont {J.~D.}\ \bibnamefont
  {Clarke}}, \bibinfo {author} {\bibfnamefont {R.}~\bibnamefont {Foot}}, \ and\
  \bibinfo {author} {\bibfnamefont {R.~R.}\ \bibnamefont {Volkas}},\ }\href
  {\doibase 10.1007/JHEP02(2014)123} {\bibfield  {journal} {\bibinfo  {journal}
  {JHEP}\ }\textbf {\bibinfo {volume} {02}},\ \bibinfo {pages} {123} (\bibinfo
  {year} {2014})},\ \Eprint {http://arxiv.org/abs/1310.8042} {arXiv:1310.8042
  [hep-ph]} \BibitemShut {NoStop}%
\bibitem [{\citenamefont {Monin}\ \emph {et~al.}(2019)\citenamefont {Monin},
  \citenamefont {Boyarsky},\ and\ \citenamefont {Ruchayskiy}}]{Monin:2018lee}%
  \BibitemOpen
  \bibfield  {author} {\bibinfo {author} {\bibfnamefont {A.}~\bibnamefont
  {Monin}}, \bibinfo {author} {\bibfnamefont {A.}~\bibnamefont {Boyarsky}}, \
  and\ \bibinfo {author} {\bibfnamefont {O.}~\bibnamefont {Ruchayskiy}},\
  }\href {\doibase 10.1103/PhysRevD.99.015019} {\bibfield  {journal} {\bibinfo
  {journal} {Phys. Rev. D}\ }\textbf {\bibinfo {volume} {99}},\ \bibinfo
  {pages} {015019} (\bibinfo {year} {2019})},\ \Eprint
  {http://arxiv.org/abs/1806.07759} {arXiv:1806.07759 [hep-ph]} \BibitemShut
  {NoStop}%
\bibitem [{\citenamefont {Bezrukov}\ \emph {et~al.}(2018)\citenamefont
  {Bezrukov}, \citenamefont {Gorbunov},\ and\ \citenamefont
  {Timiryasov}}]{Bezrukov:2018yvd}%
  \BibitemOpen
  \bibfield  {author} {\bibinfo {author} {\bibfnamefont {F.}~\bibnamefont
  {Bezrukov}}, \bibinfo {author} {\bibfnamefont {D.}~\bibnamefont {Gorbunov}},
  \ and\ \bibinfo {author} {\bibfnamefont {I.}~\bibnamefont {Timiryasov}},\
  }\href@noop {} {\  (\bibinfo {year} {2018})},\ \Eprint
  {http://arxiv.org/abs/1812.08088} {arXiv:1812.08088 [hep-ph]} \BibitemShut
  {NoStop}%
\bibitem [{\citenamefont {Diehl}\ \emph {et~al.}(1998)\citenamefont {Diehl},
  \citenamefont {Gousset}, \citenamefont {Pire},\ and\ \citenamefont
  {Teryaev}}]{Diehl:1998dk}%
  \BibitemOpen
  \bibfield  {author} {\bibinfo {author} {\bibfnamefont {M.}~\bibnamefont
  {Diehl}}, \bibinfo {author} {\bibfnamefont {T.}~\bibnamefont {Gousset}},
  \bibinfo {author} {\bibfnamefont {B.}~\bibnamefont {Pire}}, \ and\ \bibinfo
  {author} {\bibfnamefont {O.}~\bibnamefont {Teryaev}},\ }\href {\doibase
  10.1103/PhysRevLett.81.1782} {\bibfield  {journal} {\bibinfo  {journal}
  {Phys. Rev. Lett.}\ }\textbf {\bibinfo {volume} {81}},\ \bibinfo {pages}
  {1782} (\bibinfo {year} {1998})},\ \Eprint
  {http://arxiv.org/abs/hep-ph/9805380} {arXiv:hep-ph/9805380} \BibitemShut
  {NoStop}%
\bibitem [{\citenamefont {Diehl}(2003)}]{Diehl:2003ny}%
  \BibitemOpen
  \bibfield  {author} {\bibinfo {author} {\bibfnamefont {M.}~\bibnamefont
  {Diehl}},\ }\href {\doibase 10.1016/j.physrep.2003.08.002} {\bibfield
  {journal} {\bibinfo  {journal} {Phys. Rept.}\ }\textbf {\bibinfo {volume}
  {388}},\ \bibinfo {pages} {41} (\bibinfo {year} {2003})},\ \Eprint
  {http://arxiv.org/abs/hep-ph/0307382} {arXiv:hep-ph/0307382} \BibitemShut
  {NoStop}%
\bibitem [{\citenamefont {Soffer}\ and\ \citenamefont
  {Teryaev}(2004)}]{Soffer:2004ip}%
  \BibitemOpen
  \bibfield  {author} {\bibinfo {author} {\bibfnamefont {J.}~\bibnamefont
  {Soffer}}\ and\ \bibinfo {author} {\bibfnamefont {O.}~\bibnamefont
  {Teryaev}},\ }\href {\doibase 10.1103/PhysRevD.70.116004} {\bibfield
  {journal} {\bibinfo  {journal} {Phys. Rev. D}\ }\textbf {\bibinfo {volume}
  {70}},\ \bibinfo {pages} {116004} (\bibinfo {year} {2004})},\ \Eprint
  {http://arxiv.org/abs/hep-ph/0410228} {arXiv:hep-ph/0410228} \BibitemShut
  {NoStop}%
\bibitem [{\citenamefont {Vainshtein}\ \emph {et~al.}(1980)\citenamefont
  {Vainshtein}, \citenamefont {Zakharov},\ and\ \citenamefont
  {Shifman}}]{Vainshtein:1980ea}%
  \BibitemOpen
  \bibfield  {author} {\bibinfo {author} {\bibfnamefont {A.~I.}\ \bibnamefont
  {Vainshtein}}, \bibinfo {author} {\bibfnamefont {V.~I.}\ \bibnamefont
  {Zakharov}}, \ and\ \bibinfo {author} {\bibfnamefont {M.~A.}\ \bibnamefont
  {Shifman}},\ }\href {\doibase 10.1070/PU1980v023n08ABEH005019} {\bibfield
  {journal} {\bibinfo  {journal} {Sov. Phys. Usp.}\ }\textbf {\bibinfo {volume}
  {23}},\ \bibinfo {pages} {429} (\bibinfo {year} {1980})}\BibitemShut
  {NoStop}%
\bibitem [{\citenamefont {Voloshin}(1986)}]{Voloshin:1985tc}%
  \BibitemOpen
  \bibfield  {author} {\bibinfo {author} {\bibfnamefont {M.~B.}\ \bibnamefont
  {Voloshin}},\ }\href@noop {} {\bibfield  {journal} {\bibinfo  {journal} {Sov.
  J. Nucl. Phys.}\ }\textbf {\bibinfo {volume} {44}},\ \bibinfo {pages} {478}
  (\bibinfo {year} {1986})}\BibitemShut {NoStop}%
\bibitem [{\citenamefont {Burkert}\ \emph {et~al.}(2023)\citenamefont
  {Burkert}, \citenamefont {Elouadrhiri}, \citenamefont {Girod}, \citenamefont
  {Lorce}, \citenamefont {Schweitzer},\ and\ \citenamefont
  {Shanahan}}]{Burkert:2023wzr}%
  \BibitemOpen
  \bibfield  {author} {\bibinfo {author} {\bibfnamefont {V.~D.}\ \bibnamefont
  {Burkert}}, \bibinfo {author} {\bibfnamefont {L.}~\bibnamefont
  {Elouadrhiri}}, \bibinfo {author} {\bibfnamefont {F.~X.}\ \bibnamefont
  {Girod}}, \bibinfo {author} {\bibfnamefont {C.}~\bibnamefont {Lorce}},
  \bibinfo {author} {\bibfnamefont {P.}~\bibnamefont {Schweitzer}}, \ and\
  \bibinfo {author} {\bibfnamefont {P.~E.}\ \bibnamefont {Shanahan}},\
  }\href@noop {} {\  (\bibinfo {year} {2023})},\ \Eprint
  {http://arxiv.org/abs/2303.08347} {arXiv:2303.08347 [hep-ph]} \BibitemShut
  {NoStop}%
\bibitem [{\citenamefont {Teryaev}(2016)}]{Teryaev:2016edw}%
  \BibitemOpen
  \bibfield  {author} {\bibinfo {author} {\bibfnamefont {O.~V.}\ \bibnamefont
  {Teryaev}},\ }\href {\doibase 10.1007/s11467-016-0573-6} {\bibfield
  {journal} {\bibinfo  {journal} {Front. Phys. (Beijing)}\ }\textbf {\bibinfo
  {volume} {11}},\ \bibinfo {pages} {111207} (\bibinfo {year}
  {2016})}\BibitemShut {NoStop}%
\bibitem [{\citenamefont {Kumano}\ \emph {et~al.}(2018)\citenamefont {Kumano},
  \citenamefont {Song},\ and\ \citenamefont {Teryaev}}]{Kumano:2017lhr}%
  \BibitemOpen
  \bibfield  {author} {\bibinfo {author} {\bibfnamefont {S.}~\bibnamefont
  {Kumano}}, \bibinfo {author} {\bibfnamefont {Q.-T.}\ \bibnamefont {Song}}, \
  and\ \bibinfo {author} {\bibfnamefont {O.~V.}\ \bibnamefont {Teryaev}},\
  }\href {\doibase 10.1103/PhysRevD.97.014020} {\bibfield  {journal} {\bibinfo
  {journal} {Phys. Rev. D}\ }\textbf {\bibinfo {volume} {97}},\ \bibinfo
  {pages} {014020} (\bibinfo {year} {2018})},\ \Eprint
  {http://arxiv.org/abs/1711.08088} {arXiv:1711.08088 [hep-ph]} \BibitemShut
  {NoStop}%
\bibitem [{\citenamefont {Byd\v{z}ovsk\'y}\ \emph {et~al.}(2016)\citenamefont
  {Byd\v{z}ovsk\'y}, \citenamefont {Kami\'nski},\ and\ \citenamefont
  {Nazari}}]{Bydzovsky:2016vdx}%
  \BibitemOpen
  \bibfield  {author} {\bibinfo {author} {\bibfnamefont {P.}~\bibnamefont
  {Byd\v{z}ovsk\'y}}, \bibinfo {author} {\bibfnamefont {R.}~\bibnamefont
  {Kami\'nski}}, \ and\ \bibinfo {author} {\bibfnamefont {V.}~\bibnamefont
  {Nazari}},\ }\href {\doibase 10.1103/PhysRevD.94.116013} {\bibfield
  {journal} {\bibinfo  {journal} {Phys. Rev. D}\ }\textbf {\bibinfo {volume}
  {94}},\ \bibinfo {pages} {116013} (\bibinfo {year} {2016})},\ \Eprint
  {http://arxiv.org/abs/1611.10070} {arXiv:1611.10070 [hep-ph]} \BibitemShut
  {NoStop}%
\bibitem [{\citenamefont {Gasser}\ and\ \citenamefont
  {Leutwyler}(1984)}]{Gasser:1983yg}%
  \BibitemOpen
  \bibfield  {author} {\bibinfo {author} {\bibfnamefont {J.}~\bibnamefont
  {Gasser}}\ and\ \bibinfo {author} {\bibfnamefont {H.}~\bibnamefont
  {Leutwyler}},\ }\href {\doibase 10.1016/0003-4916(84)90242-2} {\bibfield
  {journal} {\bibinfo  {journal} {Annals Phys.}\ }\textbf {\bibinfo {volume}
  {158}},\ \bibinfo {pages} {142} (\bibinfo {year} {1984})}\BibitemShut
  {NoStop}%
\bibitem [{\citenamefont {Kubis}\ and\ \citenamefont
  {Meissner}(2000)}]{Kubis:1999db}%
  \BibitemOpen
  \bibfield  {author} {\bibinfo {author} {\bibfnamefont {B.}~\bibnamefont
  {Kubis}}\ and\ \bibinfo {author} {\bibfnamefont {U.-G.}\ \bibnamefont
  {Meissner}},\ }\href {\doibase 10.1016/S0375-9474(99)00823-4} {\bibfield
  {journal} {\bibinfo  {journal} {Nucl. Phys. A}\ }\textbf {\bibinfo {volume}
  {671}},\ \bibinfo {pages} {332} (\bibinfo {year} {2000})},\ \bibinfo {note}
  {[Erratum: Nucl.Phys.A 692, 647--648 (2001)]},\ \Eprint
  {http://arxiv.org/abs/hep-ph/9908261} {arXiv:hep-ph/9908261} \BibitemShut
  {NoStop}%
\bibitem [{\citenamefont {Djouadi}\ \emph {et~al.}(1991)\citenamefont
  {Djouadi}, \citenamefont {Spira},\ and\ \citenamefont
  {Zerwas}}]{Djouadi:1991tka}%
  \BibitemOpen
  \bibfield  {author} {\bibinfo {author} {\bibfnamefont {A.}~\bibnamefont
  {Djouadi}}, \bibinfo {author} {\bibfnamefont {M.}~\bibnamefont {Spira}}, \
  and\ \bibinfo {author} {\bibfnamefont {P.~M.}\ \bibnamefont {Zerwas}},\
  }\href {\doibase 10.1016/0370-2693(91)90375-Z} {\bibfield  {journal}
  {\bibinfo  {journal} {Phys. Lett. B}\ }\textbf {\bibinfo {volume} {264}},\
  \bibinfo {pages} {440} (\bibinfo {year} {1991})}\BibitemShut {NoStop}%
\bibitem [{\citenamefont {Lorc\'e}\ \emph {et~al.}(2022)\citenamefont
  {Lorc\'e}, \citenamefont {Pire},\ and\ \citenamefont {Song}}]{Lorce:2022tiq}%
  \BibitemOpen
  \bibfield  {author} {\bibinfo {author} {\bibfnamefont {C.}~\bibnamefont
  {Lorc\'e}}, \bibinfo {author} {\bibfnamefont {B.}~\bibnamefont {Pire}}, \
  and\ \bibinfo {author} {\bibfnamefont {Q.-T.}\ \bibnamefont {Song}},\ }\href
  {\doibase 10.1103/PhysRevD.106.094030} {\bibfield  {journal} {\bibinfo
  {journal} {Phys. Rev. D}\ }\textbf {\bibinfo {volume} {106}},\ \bibinfo
  {pages} {094030} (\bibinfo {year} {2022})},\ \Eprint
  {http://arxiv.org/abs/2209.11140} {arXiv:2209.11140 [hep-ph]} \BibitemShut
  {NoStop}%
\bibitem [{\citenamefont {Colangelo}(2002)}]{Colangelo:2001uv}%
  \BibitemOpen
  \bibfield  {author} {\bibinfo {author} {\bibfnamefont {G.}~\bibnamefont
  {Colangelo}},\ }\href {\doibase 10.1016/S0920-5632(01)01643-7} {\bibfield
  {journal} {\bibinfo  {journal} {Nucl. Phys. B Proc. Suppl.}\ }\textbf
  {\bibinfo {volume} {106}},\ \bibinfo {pages} {53} (\bibinfo {year} {2002})},\
  \Eprint {http://arxiv.org/abs/hep-lat/0111003} {arXiv:hep-lat/0111003}
  \BibitemShut {NoStop}%
\bibitem [{\citenamefont {Achasov}\ \emph {et~al.}(2021)\citenamefont
  {Achasov}, \citenamefont {Bennett}, \citenamefont {Kiselev}, \citenamefont
  {Kozyrev},\ and\ \citenamefont {Shestakov}}]{Achasov:2020aun}%
  \BibitemOpen
  \bibfield  {author} {\bibinfo {author} {\bibfnamefont {N.~N.}\ \bibnamefont
  {Achasov}}, \bibinfo {author} {\bibfnamefont {J.~V.}\ \bibnamefont
  {Bennett}}, \bibinfo {author} {\bibfnamefont {A.~V.}\ \bibnamefont
  {Kiselev}}, \bibinfo {author} {\bibfnamefont {E.~A.}\ \bibnamefont
  {Kozyrev}}, \ and\ \bibinfo {author} {\bibfnamefont {G.~N.}\ \bibnamefont
  {Shestakov}},\ }\href {\doibase 10.1103/PhysRevD.103.014010} {\bibfield
  {journal} {\bibinfo  {journal} {Phys. Rev. D}\ }\textbf {\bibinfo {volume}
  {103}},\ \bibinfo {pages} {014010} (\bibinfo {year} {2021})},\ \Eprint
  {http://arxiv.org/abs/2009.04191} {arXiv:2009.04191 [hep-ph]} \BibitemShut
  {NoStop}%
\bibitem [{\citenamefont {Workman}\ \emph {et~al.}(2022)\citenamefont {Workman}
  \emph {et~al.}}]{ParticleDataGroup:2022pth}%
  \BibitemOpen
  \bibfield  {author} {\bibinfo {author} {\bibfnamefont {R.~L.}\ \bibnamefont
  {Workman}} \emph {et~al.} (\bibinfo {collaboration} {Particle Data Group}),\
  }\href {\doibase 10.1093/ptep/ptac097} {\bibfield  {journal} {\bibinfo
  {journal} {PTEP}\ }\textbf {\bibinfo {volume} {2022}},\ \bibinfo {pages}
  {083C01} (\bibinfo {year} {2022})}\BibitemShut {NoStop}%
\bibitem [{\citenamefont {Anikin}\ \emph {et~al.}(2005)\citenamefont {Anikin},
  \citenamefont {Pire},\ and\ \citenamefont {Teryaev}}]{Anikin:2005ur}%
  \BibitemOpen
  \bibfield  {author} {\bibinfo {author} {\bibfnamefont {I.~V.}\ \bibnamefont
  {Anikin}}, \bibinfo {author} {\bibfnamefont {B.}~\bibnamefont {Pire}}, \ and\
  \bibinfo {author} {\bibfnamefont {O.~V.}\ \bibnamefont {Teryaev}},\ }\href
  {\doibase 10.1016/j.physletb.2005.08.113} {\bibfield  {journal} {\bibinfo
  {journal} {Phys. Lett. B}\ }\textbf {\bibinfo {volume} {626}},\ \bibinfo
  {pages} {86} (\bibinfo {year} {2005})},\ \Eprint
  {http://arxiv.org/abs/hep-ph/0506277} {arXiv:hep-ph/0506277} \BibitemShut
  {NoStop}%
\bibitem [{\citenamefont {Achasov}\ \emph {et~al.}(1982)\citenamefont
  {Achasov}, \citenamefont {Devyanin},\ and\ \citenamefont
  {Shestakov}}]{Achasov:1982bt}%
  \BibitemOpen
  \bibfield  {author} {\bibinfo {author} {\bibfnamefont {N.~N.}\ \bibnamefont
  {Achasov}}, \bibinfo {author} {\bibfnamefont {S.~A.}\ \bibnamefont
  {Devyanin}}, \ and\ \bibinfo {author} {\bibfnamefont {G.~N.}\ \bibnamefont
  {Shestakov}},\ }\href {\doibase 10.1007/BF01573747} {\bibfield  {journal}
  {\bibinfo  {journal} {Z. Phys. C}\ }\textbf {\bibinfo {volume} {16}},\
  \bibinfo {pages} {55} (\bibinfo {year} {1982})}\BibitemShut {NoStop}%
\bibitem [{\citenamefont {Barry}\ \emph {et~al.}(2018)\citenamefont {Barry},
  \citenamefont {Sato}, \citenamefont {Melnitchouk},\ and\ \citenamefont
  {Ji}}]{Barry:2018ort}%
  \BibitemOpen
  \bibfield  {author} {\bibinfo {author} {\bibfnamefont {P.~C.}\ \bibnamefont
  {Barry}}, \bibinfo {author} {\bibfnamefont {N.}~\bibnamefont {Sato}},
  \bibinfo {author} {\bibfnamefont {W.}~\bibnamefont {Melnitchouk}}, \ and\
  \bibinfo {author} {\bibfnamefont {C.-R.}\ \bibnamefont {Ji}},\ }\href
  {\doibase 10.1103/PhysRevLett.121.152001} {\bibfield  {journal} {\bibinfo
  {journal} {Phys. Rev. Lett.}\ }\textbf {\bibinfo {volume} {121}},\ \bibinfo
  {pages} {152001} (\bibinfo {year} {2018})},\ \Eprint
  {http://arxiv.org/abs/1804.01965} {arXiv:1804.01965 [hep-ph]} \BibitemShut
  {NoStop}%
\end{thebibliography}%

\end{document}